\documentclass[pra,aps,twocolumn,showpacs,epsfig]{revtex4}
\usepackage{color}
\usepackage{graphicx,floatflt,amssymb,epsfig,epsf,amsmath} 

\begin{document}

\title{Supersymmetric isospectral formalism for the calculation of near-zero energy states: application to the very weakly bound ${^4}$He trimer excited state} 

\author{Sudip Kumar Haldar$^{1}$\footnote{e-mail: sudip\_cu@rediffmail.com}, Barnali Chakrabarti$^{1,2}$, Tapan Kumar Das$^{3}$} 

\affiliation{
$^{1}$Department of Physics, Lady Brabourne College, 
P-1/2 Suhrawardi Avenue, Kolkata 700017, India.\\
$^{2}$Instituto de Fisica, Universidade de S\~ao Paulo, CP 66318, 05315-970, S\~ao Paulo, SP Brazil\\
$^{3}$Department of Physics, University of Calcutta, 92 A P C Road, Kolkata 700009, India}
\begin{abstract}
We propose a novel mathematical approach for the calculation of near-zero energy states by solving potentials which are isospectral 
with the original one. For any potential, families of strictly isospectral potentials (with very different 
shape) having desirable and adjustable features are generated by supersymmetric isospectral formalism. The near-zero energy Efimov 
state in the original potential is effectively trapped in the deep well of the isospectral 
family and facilitates more accurate calculation of the Efimov state. Application to the first excited state in $^{4}$He  trimer is presented.
\end{abstract}
\pacs{03.75.Hh, 31.15.Ja, 03.65.Ge, 03.75.Nt}
\maketitle

\section{Introduction}
It was proposed by Efimov in 1970 that if two spinless neutral 
bosons interact resonantly then the addition of a third 
identical particle leads to the appearance of an infinite number 
of bound three-body energy levels~\cite{uhi,iju}. This occurs simultaneously 
with the divergence of the $s$-wave 
scattering length $(a_{s})$, associated with appearance of an 
additional zero-energy two-body bound state. Hence highly 
exotic Efimov states appear when there is a zero or near-zero 
energy two-body bound state. For a long time there was no 
clear signature of 
Efimov states in any naturally occuring trimer system. Efimov 
states are not possible in atomic systems due to the long 
range Coulomb interaction, however it may exist in the system 
of spinless neutral atoms. Even though the Efimov effect was 
predicted four decades ago~\cite{uhi,iju}, evidence of its existence 
in ultracold caesium and potassium trimers has been 
experimentally established only very recently~\cite{KKZ}. However, 
these trimers are obtained by manipulating two-body forces 
through Feshbach resonances and are not naturally occuring. 
Therefore, it is of great interest 
to search for the Efimov effect in a naturally occuring 
trimer, like $^4$He trimer. So far no experimental 
confirmation has been reported. The near-zero energy 
($E_{0} \sim$ 1 mK) bound 
state (which is also the only bound state) of $^{4}$He dimer 
opens the possibility of the existence of an Efimov-like state 
in $^{4}$He trimer. Several authors remarked that the $^{4}$He 
trimer may be the most promising candidate. Earlier theoretical 
calculations show that the trimer has a $L$ =0 ground state 
at 126 mK and a near-zero energy excited state ($\sim$ 2mK)~\cite{gft,tgy,esw,sad,yhu,ijj,uuu,szx,swa,gff}. 
The excited state has been claimed to be an Efimov state. 
A controversy arises from the fact that the number of Efimov 
states is highly sensitive to the binding energy of the dimer 
and even a very small decrease of the strength of two-body 
interaction makes the system unbound. Strikingly, it also disappears 
when the two-body interaction strength is {\it increased}. 
However in contrast with theoretical investigations, no evidence of Efimov 
trimer has been found experimentally~\cite{ews,ytg}. In the experiments, 
$^{4}$He trimer has been observed in its ground state only. 
No experimental evidence of the excited state has been reported 
so far. \\

In principle $^{4}$He trimer may be considered as a very simple 
three-body system consisting of three identical 
bosons. But its theoretical treatment is quite difficult. 
First, the He-dimer potential is not uniquely known. Very small 
uncertainities in the dimer 
potential may lead to different conclusions. Secondly, the strong 
short-range interatomic repulsion in the He-He interaction causes 
large numerical errors. 
As $^{4}$He systems are strongly correlated due to large 
$^{4}$He--$^{4}$He repulsion at short separation, the effect of 
interatomic correlation must be taken properly into account. \\

In the present communication, we revisit the problem using a 
correlated basis function known as potential harmonics (PH) basis 
which takes care of two-body 
correlations~\cite{fabre}. In order to include the effect of highly 
repulsive He-He core, we multiply the PH basis with a suitable 
short-range correlation function which 
reproduces the correct short-range behavior of the dimer wavefunction. 
Although this correlated PH  basis (CPH basis) correctly 
reproduces the dimer and trimer properties, we could not find any 
Efimov like state in trimer with the actual dimer interaction~\cite{hhh}. 
We point out that the calculation of such a 
near-zero energy excited state in the shallow and extended trimer 
potential may involve severe numerical errors and we may miss it. 
Thus an alternative accurate procedure is desirable. Here, we apply 
the supersymmetric isospectral formalism for an accurate treatment. 
For any given potential, families of strictly isospectral potentials, 
{\it with very different shape} but having desirable and adjustable features 
are generated by supersymmetric isospectral formalism~\cite{ygt}. The 
near-zero energy bound state will be more effectively bound in the 
deep narrow well of the isospectral potential and 
will facilitate an easier and more accurate calculation of the near-zero 
energy excited state. 
Following the steps of supersymmetric quantum mechanics ~\cite{ygt}, 
for any given potential $\omega_0(r)$, one can construct a class of 
potentials ${\hat{\omega}}_{0}(r,\{\lambda\})$, where 
$\{\lambda\}$ represents a set of one or more continuously variable 
real parameters. The potential ${\hat{\omega}}_{0}$ is isospectral in 
the sense that $\omega_{0}$ and ${\hat{\omega}}_{0}$ have identical spectrum, 
reflection and tranmission coefficients. For simplicity we consider 
only one parameter $(\lambda)$ family of isospectral potentials. 
We will see later that $\lambda$ can take real values 
$-\infty < \lambda < -1$ and $ 0<\lambda<\infty $. For 
$\lambda \rightarrow \infty$, one gets back 
the original potential. Although the set of isospectral potentials 
are strictly isospectral with the original potential, they have 
different shapes depending on the parameter $\lambda$~\cite{ygt}. 
\begin{figure}
\noindent \includegraphics[clip,width= 9cm,height=9cm,angle=0    
]{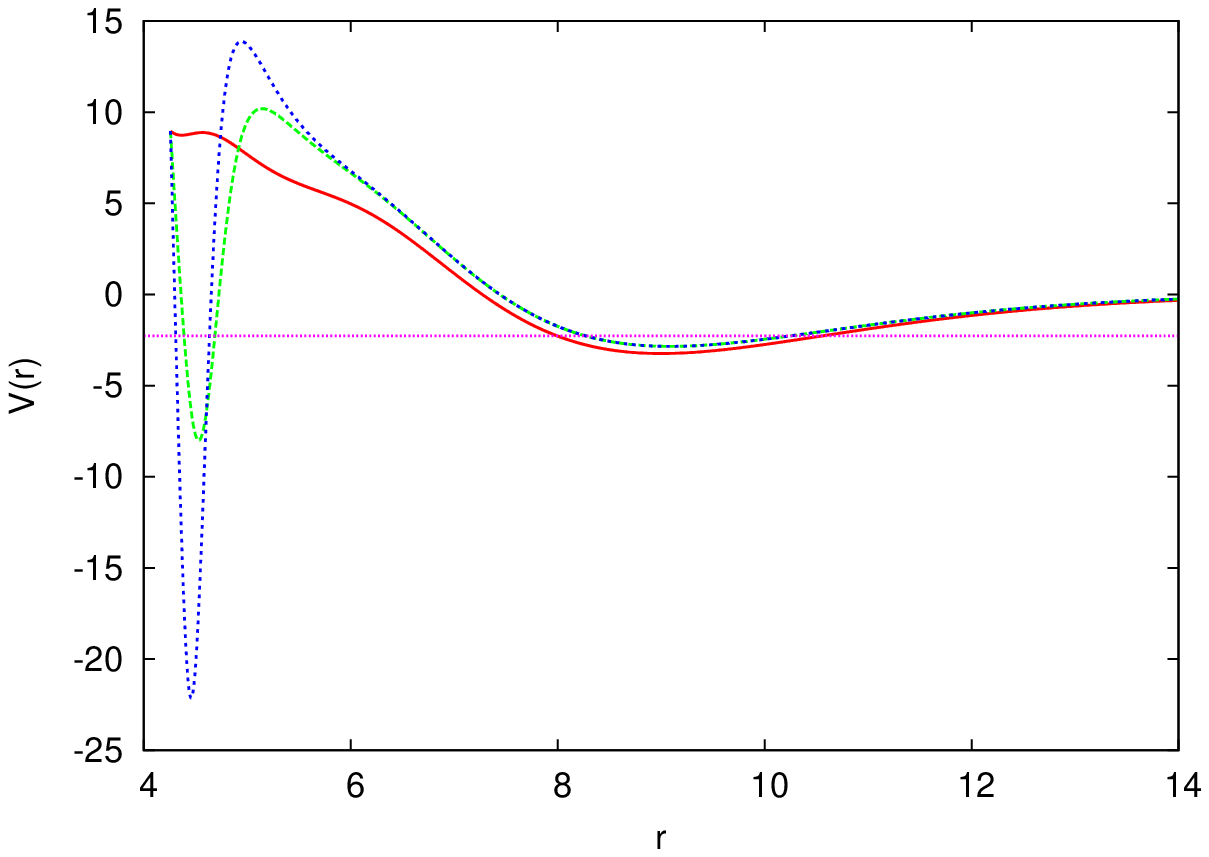}\\                                                           
Fig. 1 (Color online) The effective potential $\omega_0(r)$ (red solid curve) and isospectral 
potentials $\hat{\omega}_0(r,\lambda)$ corresponding to two values of $\lambda$: 
$\lambda=0.00005$ (green 
dashed curve) and $\lambda=0.00002$ (blue dotted curve) for the $^4$He trimer. 
All energies are in mK and $r$ in a.u. The horizontal line indicates the energy 
of the first excited state in $\omega_0(r)$. 
\end{figure}
In Fig.~1, we demonstrate how an original potential, $\omega_0(r)$ 
shown by the solid (red) curve, changes in the isospectral 
potential $\hat{\omega}_0(r,\lambda)$ for two values of the 
parameter $\lambda$, {\it viz.} $\lambda=0.00005$ (green dashed curve) 
and $\lambda=0.00002$ (blue dotted curve). We introduce this figure 
here for a qualitative understanding of the features of the isospectral 
potentials. A complete discussion of how such isospectral potentials 
are calculated will be presented in Sections IIC and III. Although all three 
potentials produce identical energy spectrum, their shapes are seen to be 
very different. The original potential [$\omega_0(r)$] has a shallow 
and wide well with a short range repulsion. By contrast, both the 
isospectral potentials have a deep and narrow {\it attractive} well (NAW) at 
smaller $r$, while the long range part does not differ much from 
$\omega_0(r)$. One can also notice that as $\lambda$ decreases, the 
narrow attractive well becomes deeper, while the intermediate barrier 
becomes higher. Hence, the near-zero energy (indicated in Fig.~1 by a 
horizontal line) excited state in $\omega_0(r)$ 
will be at the {\it same energy value} in the isospectral potentials, 
but now that state will lie {\it deep within the NAW}. 
Thus, while this state is weakly bound and spatially extended in 
the original potential, it will be strongly bound and well localized 
in the NAW of the isospectral potential, at the same 
energy. Clearly, computation of the 
wave function and its energy (equal to the energy of the first excited 
state in the original potential) will be easier. Furthermore, this state 
becomes more strongly bound, as $\lambda$ decreases. 
In general, as $\lambda$ decreases continuously from $\infty$ to $0$, 
${\hat{\omega}}_{0}$ starts developing a local minimum which shifts 
towards $r$=0 and becomes deeper and narrower. Consequently, a 
shallow potential transforms into one having a narrow and deep potential 
well near the origin in the isospectral potential. The surface barrier 
also becomes high. Such interesting properties of ${\hat{\omega}}_{0}$ can 
be useful to solve near-zero energy states. Such a state lies near 
the top of the original potential well and its wave function is 
spatially very extended, while the isospectral state lies 
well within the NAW. Hence the 
latter is strongly bound and well localized within the narrow well 
of $\hat{V}_1$. The parameter $\lambda$ controls these features 
and a suitable optimum value can be chosen (see later). \\

Thus our approach consists of two steps. First, to apply a correlated 
quantum many-body theory 
for a highly correlated system like $^{4}$He-trimer and second, to use 
the isospectral formalism for the accurate 
determination of the first excited state, whose energy is just a few mK. \\

The paper is organized as follows. In Section~II, we present a brief review of 
the correlated potential harmonics expansion method, choice of potential and the isospectral formalism. 
Section~III presents the results of our numerical calculation. Finally we 
draw our conclusions in Section~IV. \\

\section{Theoretical procedure}

\subsection{Correlated Potential harmonics expansion method}
The Hamiltonian for a system of $(N+1)$ atoms ( each of mass $m$) and interacting via 
two-body potential has the form
\begin{equation}
H = - \frac{\hbar^{2}}{2m} \sum_{i=1}^{N+1}\nabla_{i}^{2}+ \sum_{i>j=1}^{N+1}V(\vec{x}_{i}-\vec{x}_{j})
\end{equation}
where $V(\vec{x}_{i}-\vec{x}_{j})$ = $V(\vec{r}_{ij})$ is the He-He two-body potential described later. 
The relative motion in the standard Jacobi coordinates, $\{ \vec{\zeta}_1, \dots, \vec{\zeta}_N\}$, is given by~\cite{fabre}
\begin{equation}
\left[-\frac{\hbar^{2}}{m}\sum_{i=1}^{N}\nabla_{\zeta_{i}}^{2}+
V(\vec{\zeta}_1, \dots, \vec{\zeta}_N)
-E \right]
\Psi(\vec{\zeta}_1,\dots,\vec{\zeta}_N)=0. 
\end{equation} 
We decompose $\Psi$ in Faddeev components
\begin{equation}
\Psi(\vec{x}) = \sum_{ij>i}^{N+1} \psi_{ij}(\vec{x}),
\label{PHexpn}
\end{equation}
where $\psi_{ij}(\vec{x})$ is the two-body Faddeev component for the $(ij)$ partition. In potential harmonics expansion method (PHEM), we expand $(ij)$-Faddeev component in the complete set of potential harmonics, 
$\{{\mathcal P}_{2K+l}^{lm}(\Omega_{N}^{(ij)})\}$, 
appropriate for $(ij)$-partition~\cite{fabre,rft} 
\begin{equation}
\psi_{ij}
=r^{-(\frac{3N-1}{2})}\sum_{K}{\mathcal P}_{2K+l}^{lm}
(\Omega_{N}^{(ij)})u_{K}^{l}(r).
\end{equation}
PH basis is the subset of hyperspherical harmonics 
(HH), which is sufficient 
for the expansion of the two-body potential $V(\vec{r}_{ij})$, for the 
particular $(ij)$ partition~\cite{fabre}. This PH function depends only on 
$\vec{r}_{ij}$ and a global length $r$ (called hyperradius). Hence 
$\Psi$ in Eq.~(\ref{PHexpn}) 
includes only long range two-body correlations. $\Omega_{N}^{(ij)}$ represents the full set of hyperangles for the $(ij)$ partition. 
The potential harmonics basis function for the $(ij)$-partition 
is given by~[16]
\begin{equation}
P_{2K+l}^{lm}(\Omega_{ij}) = Y_{lm}(\omega_{ij}) ^{(N)}P_{2K+l}^{l,0}(\phi)\mathcal {Y}_{0}(D-3),
\end{equation} 
where $\phi=\cos^{-1}(r_{ij}/r)$, $Y_{lm}(\omega_{ij})$ 
is a spherical harmonic, $\omega_{ij}$ being the polar angles 
of $\vec{r}_{ij}$,
$^{(N)}P_{2K+l}^{l,0}(\phi)$ is expressed in terms of Jacobi polynomials and $\mathcal {Y}_{0}$ is the lowest order hyperspherical harmonic in $(3N-3)$ 
dimensional hyperangular space (hence a constant)~\cite{rft}. The basic 
assumption in Eq.~(3) is that only two-body correlations are 
important and higher-body correlations are disregarded. 
Thus in the $(ij)$-Faddeev component, where only the $(ij)$-pair interacts, $\psi_{ij}$ is independent of the coordinates of all particles except $\vec{r}_{ij}$.  Hence we can freeze the contributions coming from $(N-1)$ remaining spectators. Thus the contribution to the orbital angular momentum and the grand orbital quantum 
number comes only from the interacting pair and the $3N$ dimensional Schr\"odinger equation reduces effectively to a four dimensional equation. 
The relevant set of quantum numbers, associated with the hyperangles, are three. These are orbital $l$, azimuthal $m$ and grand orbital $2K+l$ for any $N$. \\

So far we have disregarded the effect of strong short range correlation in the PH basis. The He-He potential becomes suddenly very 
strongly repulsive below a certain value of interatomic separation. This causes a very strong short range two-body correlation 
in the many body wave function. We introduce this correlation function in the expansion basis and 
call it as CPH basis~\cite{edd}. 
\begin{eqnarray}
\left[{\mathcal P}_{2K+l}^{l,m} (\Omega_{(ij)})\right]_{correlated} =
Y_{lm}(\omega_{ij})\hspace*{.1cm} 
^{(N)} P_{2K+l}^{l,0}(\phi)
\nonumber
\\
 {\mathcal Y}_{0}(3N-3) \eta(r_{ij}) ,~~~~~~
\end{eqnarray}
where $\eta(r_{ij})$ is the short range correlation function. It is practically zero for very small $r_{ij}$ $(r_{ij}<r_{c})$ where $r_{c}$ 
is the size of the repulsive core. Its role is to enhance the speed of convergence of the expansion basis. We obtain
$\eta(r_{ij})$ as the zero energy solution of $(ij)$-pair relative motion in the potential $V({r}_{ij})$,
\begin{equation}
-\frac{\hbar^2}{m}\frac{1}{r_{ij}^2}\frac{d}{dr_{ij}}\left(r_{ij}^2
\frac{d\eta(r_{ij})}{dr_{ij}}\right)+V(r_{ij})\eta(r_{ij})=0  .
\end{equation}
Replacing PH by CPH in Eq.~(4) and taking projection of the Schr\"odinger 
equation on the PH basis of a particular partition, a set of coupled 
differential equation (CDE) is obtained~\cite{edd}
\begin{eqnarray}
\Big[&-&\dfrac{\hbar^{2}}{m} \dfrac{d^{2}}{dr^{2}} + \dfrac{\hbar^{2}}{mr^{2}}
\{ {\cal L}({\cal L}+1) + 4K(K+\alpha+\beta+1)\} 
\nonumber 
\\
&-& E  \Big] U_{Kl}(r) 
\\
&+& \sum_{K^{\prime}}f_{Kl}V_{KK^{\prime}}(r)f_{K'l} U_{K^{\prime}l}(r) = 0 ,
\nonumber
\end{eqnarray}
where ${{\cal L}} = l+\frac{3N-3}{2}$, $\alpha=\frac{3N-5}{2}$,
$\beta=l+\frac{1}{2}$ and $f_{Kl}$ is a constant representing the 
overlap of a PH corresponding to a particular partition with the sum 
of PHs of all partitions~\cite{edd}. $K$ is the hyperangular momentum quantum number. The correlated potential matrix element ${V}_{KK^{\prime}}(r)$ is now  
given by~\cite{edd}
\begin{eqnarray}
V_{KK^{\prime}}(r) =(h_{K}^{\alpha\beta}
h_{K^{\prime}}^{\alpha\beta})^{-\frac{1}{2}}\int_{-1}^{+1}P_{K}^{\alpha
\beta}(z) 
V\left(r\sqrt{\frac{1+z}{2}}\right)
\nonumber
\\
P_{K^{\prime}}^{\alpha \beta}(z)\eta\left(r\sqrt{\frac{1+z}{2}}\right)
W_{l}(z) dz .~~~~~~~~
\end{eqnarray}
Here $h_{K}^{\alpha\beta}$ and $W_{l}(z)$ are respectively the norm and weight function of the Jacobi 
polynomial $P_{K}^{\alpha\beta}$~\cite{fabre}.\\

\subsection{Choice of He-He potential}
For a realistic calculation, one needs an accurate He-He interaction potential. Several sophisticated He-He potentials have been proposed~\cite{arj}. Among these,the commonly used ones are: 
Tang, Tonnies and Yiu (TTY)~\cite{fdr}, 
LM2M2~\cite{raa}, 
and HFD-HE2~\cite{raaz} 
potentials. These potentials reproduce all known two-body He-He data. 
In the present work, 
we select the more popular and sophisticated TTY potential. This potential 
has the form~\cite{szx,fdr}
\begin{equation}
V(x) = A [V_{ex}(x) + V_{disp}(x)], 
\end{equation}
where $x$ represents the interparticle distance. The part $V_{ex}$ has the form $V_{ex}(x)$ = $D x^{p}e^{-2\gamma x}$ wih 
$p$ = $\frac{7}{2\gamma}-1$. The other part $V_{disp}$ is given as $V_{disp}(x)$ = $-\sum_{n=3}^{12}C_{2n}f_{2n}(x)x^{-2n}$. The 
coefficients $C_{2n}$ are calculated using the recurrence relation $C_{2n}$ = $\left(\frac{C_{2n-2}}{C_{2n-4}}\right)^{3}C_{2n-6}$;
$C_{6}$ = 1.461, $C_{8}$ = 14.11, $C_{10}$ = 183.5, $A$ = 315766.2067 $K$, D= 7.449 and $\gamma$ = 1.3443 $(a.u.)^{-1}$. The function 
$f_{2n}$ is given by $f_{2n}(x)$ = $ 1- e^{-bx}\sum_{k=0}^{2n} \frac{(bx)^{k}}{k!}$ with $b(x)$= $2\gamma$ - $\frac{p}{x}$.\\

For our numerical solution, the set of CDEs [Eq.~(8)] is solved by hyperspherical adiabatic approximation (HAA)~\cite{edr}. 
In HAA, one assumes that the hyperradial motion is slow compared to
the hyperangular motion. The effective potential for the hyperradial motion
(obtained by diagonalizing the potential matrix together with the diagonal
hypercentrifugal repulsion for a fixed value of $r$)
is obtained as a parametric function of $r$. We choose the lowest
eigenpotential ($\omega_0(r)$) as the effective potential.
Thus in HAA, energy and wavefunction are obtained approximately by solving a single uncoupled differential equation
\begin{equation}
\left[ -\frac{\hbar^{2}}{m}\frac{d^{2}}{dr^{2}} + \omega_{0}(r)-E \right]
\zeta_{0}(r)=0,
\end{equation}
subject to appropriate boundary conditions on $\zeta_{0}(r)$.
The principal advantage of the present method is two-fold.
Firstly, the CPH basis set correctly takes care of the effect of strong short range correlation produced by the He-He interaction.
Secondly, the use of HAA basically reduces the multidimensional problem to an effective one dimensional problem introducing the effective 
potential. The effective potential $(\omega_{0}(r))$ gives a clear qualitative as well as quantitative picture. For our numerical calculation we 
investigate the $l=0$ state  
and truncate the CPH basis to a maximum value $K$ = $K_{max}$, requiring proper convergence.\\

The ground state properties of $^{4}$He dimer is obtained as a numerical solution of two-body Schr\"odinger equation 
by Runga-Kutta algorithm. The dimer energy $\epsilon_{d}$ using TTY potential, as well as the results from other references are presented in Table~I.
\begin{center}
\begin{table}[!h]
\caption{The $^{4}$He-dimer energy using TTY potential.
}
\begin{tabular} {|l|lr|}
\hline
Expt. & $\epsilon_{d}$(mK)\\\cline{2-3}
$1.1_{-0.2}^{+0.3}$   & present method & -1.254 \\ \cline{2-3}
           mK        & DMC [23]      &  -1.243 \\\cline{2-3}
       [27]               & Other [11]    &  -1.309 \\\cline{2-3}
                      & Other [8]    &  -1.313 \\ \hline
\end{tabular}
\end{table}
\end{center}
Although in our earlier calculation~\cite{hhh} we reported the dimer and trimer ground state properties, we include them here for completeness  
of the discussion.
Calculated rms value of $r_{ij}$ is 
$98.596$ a.u. The extremely small binding energy of the  dimer and the 
large spatial extension of the ground state wave function imply that the 
ground state of helium dimer is very loosely bound. 
The trimer ground state energy, as well as the results obtained in earlier investigations by other authors are presented in Table~II for different 
potentials. The r.m.s. value of hyperradius 
is 21.389 $a.u.$  
Thus our correlated PH basis successfully reproduces the energy values which are in very close agreement with other 
sophisticated calculations~\cite{tgy,esw,sad,yhu,szx,trf,gho,kli,fvr}.\\

\begin{center}
\begin{table}[!h]
\caption{The absolute value of $^{4}$He-trimer ground state energy (in units of mK) 
obtained by different methods. }
\begin{tabular} {|l|l|l|l|l|l|}
\hline
Potential  & PHEM  & variational & Faddeev  & Adiabatic &  DMC \\ \hline
TTY   &  125.51  & 126.40[24]  & 126.4 [7] & - & 125.46 [25] \\ \hline
LM2M2  & 126.37  & 126.40 [25]& 126.4  [7]& 125.2 [6]& \\ \hline
HFDHE2  & 120.28  & & 117.1 [7] & 98.1 [5]& \\\hline
\end{tabular}
\end{table}
\end{center}

Although we have done detailed calculation of $^{4}$He trimer ground state energy, we failed to obtain any trimer excited state with this 
choice of two-body interaction. Very recently we have analyzed in details the behavior of $^{4}$He trimer excited states as a function of pairwise 
interaction
\begin{equation}
V_{He-He} = \delta V_{TTY}
\end{equation}
where $\delta$ controls the strength of two-body interaction. $\delta$ =1 is the physical value of the dimer interaction and $\delta$=0 corresponds 
to free particle limit when neither two-body nor three-body bound states appear. We found that by increasing $\delta$ from a small value, 
the trimer starts to support an excited state at $\delta$ =0.978~\cite{hhh}. Binding energy of the excited state gradually increases with increase in $\delta$, 
attains its maximum value at $\delta$ = 0.984, then it decreases gradually and disappears which indicates dissociation to trimer 
fragments as a dimer and monomer. Thus the disappearance of the first excited state due to both increasing and decreasing $\delta$ clearly show that the 
state is an Efimov state. But the value of $\delta$ for which this happens, does not correspond to the actual physical dimer interaction ($\delta=1$). The Efimov property of other 
excited states are also discussed in our earlier study~\cite{hhh}. However we did not claim that the trimer excited state does not exist at all, as there 
is a large body of work in this direction~\cite{gft,tgy,esw,sad,yhu,ijj,uuu,szx,swa,gff}. So this is still an ongoing issue as there is considerable controversy in the 
earlier discussions found in Refs~\cite{gft,tgy,esw,sad,yhu,ijj,uuu,szx,swa,gff}. Thus our earlier work could not resolve the problem. It may be the limitation of our basis set which is unable 
to produce such an elusive state. So it needs further study. \\

\subsection{Isospectral formalism}
For an accurate determination of the elusive near-zero energy excited state, we use the supersymmetric isospectral formalism mentioned earlier. Note that this state lies near the top of the potential well  $\omega_0(r)$ and is very extended spatially. Consequently, a convergent calculation requires a very large number of hyperspherical partial waves (corresponding to very large $K_{max}$). On the other hand, the trimer ground state lies near the bottom of this well, is well bound and localized; it therefore converges relatively easily. Now the isospectral 
potential $\hat{\omega}_0(\lambda,r)$ with sufficiently small positive $\lambda$ will have a deep well near the origin, followed by a high barrier (see below). The near-zero energy state sought after is now strongly trapped within this narrow well, {\it i.e.} it is sharply localized. Hence its convergence is achieved easily enough. The isospectral potential is obtained in terms of the trimer ground state wave function (see below),  which by the above argument is determined with a fair degree of confidence and it offers easier calculation of the excited state. \\

Since isospectral formalism is not a common topic, we briefly explain how an isospectral potential is obtained for a given potential~\cite{ygt}. 
For the given potential $\omega_0(r)$ having a normalized ground state 
$\zeta_0(r)$ with energy $E_0$,  [see Eq.~(11)], a superpotential $W(r)$ is defined as 
\begin{equation}
W(r) = -\frac{\hbar}{\sqrt{m}}\frac{\zeta_{0}^{\prime}(r)}{\zeta_{0}(r)}.
\end{equation}
Then it can easily be seen that 
\begin{equation}
\omega_0(r)-E_0=W^2(r)-\frac{\hbar}{\sqrt{m}}W^{\prime}(r),
\end{equation}
and one can define a supersymmetric partner potential $\omega^{(2)}(r)$ through
\begin{equation}
\omega^{(2)}(r)-E_0=W^2(r)+\frac{\hbar}{\sqrt{m}}W^{\prime}(r),
\end{equation}
such that $\omega_0(r)$ and its partner have the same energy spectra, except that the ground state of $\omega_0(r)$ is absent in the spectrum of its partner~\cite{ygt}. \\

Now, for the given partner potential $\omega^{(2)}(r)$, Eq.~(15) 
can be considered as a non-linear differential equation satisfied by 
the unknown function $W(r)$ (called Riccati equation). 
With this $W(r)$, we can get back $\omega_0(r)$, using Eq.~(14). 
But solution of the non-linear Eq.~(15) is not unique. For simplicity, we use 
the units in which $\frac{\hbar^2}{m}=1$. Then the most 
general solution is~\cite{ygt} 
\begin{equation}
\hat{W}(r,\lambda)=W(r)+\frac{d}{dr}\ln|I_0(r)+\lambda|,
\label{MGSP}
\end{equation}
where $\lambda$ is an integration constant and $\hat{W}(r,\lambda)$ 
is a function of $r$, parametrically dependent on $\lambda$. $I_0(r)$ 
is given by 
\begin{equation}
I_{0}(r)={\displaystyle\int}_{0}^{r}
{[\zeta_{0}(r^{\prime})]}^{2} dr^{\prime}. 
\end{equation}
Then the family of potentials ${\hat{\omega}}_{0}(r,\lambda)$ given by 
\begin{equation}
\begin{array}{rcl}
{\hat{\omega}}_{0}(r,\lambda)-E_0 &=& \hat{W}^{2}(r,\lambda)-\hat{W^{\prime}}(r,\lambda)\\
{\hat{\omega}}_{0}(r,\lambda) &=& \omega_{0}(r)-2\frac{d^{2}}{d{r}^{2}}\ln|I_{0}(r) +\lambda|
 \end{array}
\label{ISP}
 \end{equation}
for all allowed values of $\lambda$ (see below), has the {\it same 
partner} $\omega^{(2)}(r)$. Hence the family of potentials given by 
Eq.~(\ref{ISP}) all have identical spectrum. Since $\zeta_0(r)$ 
is normalized, $0 \leq I_0(r) \leq 1$. Hence from Eq.~(\ref{MGSP}), 
one notices that the interval $-1 \leq \lambda \leq 0$ is not allowed. 
For all other values of $\lambda$, $\hat{\omega}_0(r,\lambda)$ is 
strictly isospectral with $\omega_0(r)$~\cite{ygt,tfr}. 
For $\lambda=\infty$, $\hat{\omega}_0(r,\lambda)$ 
becomes $\omega_0(r)$. For small positive values of $\lambda$, the isospectral 
potential $\hat{\omega}_0(r,\lambda)$ develops an attractive well 
near the 
origin. As $\lambda \rightarrow 0$, this well becomes deeper followed 
by a high barrier, before the shallower part. 
Note that the ground state of $\omega_0(r)$, {\it viz.} $\zeta_0(r)$, is 
necessary to calculate the family of isospectral potentials. In the 
present study the ground state of $\omega_0(r)$ is fairly accurately 
calculated. We will see in the next section that 
different eigen states are transformed differently. Therefore 
this transformation cannot be considered as a generalized rotation 
in the Hilbert space.  Thus the 
transformation from the original Hamiltonian to the isospectral Hamiltonian 
is different from a standard unitary transformation, even though the 
entire energy spectrum is preserved. \\

\section{Results and discussions}

First we calculate the lowest effective potential $\omega_0(r)$ in the hyperradial space by hyperspherical adiabatic approximation (HAA) as stated in Sec.~IIB. To calculate the energy $E$ and wave function $\zeta_0(r)$ of the original potential we solve the single uncoupled equation, Eq.~(11), with appropriate boundary conditions. As stated earlier, although our calculated ground state energy $E_0$ and wave function are in good agreement with other calculations, we fail to get the first excited state in such a shallow potential. Here the isospectral formalism will be an effective technique to calculate  several isospectral potentials of gradually varying shape, which will facilitate easier calculation of the very weakly bound state. We calculate $I_0(r)$ from Eq.~(17) and then the isospectral potential for a specific value of $\lambda$ is calculated from Eq.~(18). \\

In Fig.~1, the effective 
potential $\omega_0(r)$ calculated by the CPH method is shown 
as a continuous (red) 
curve. This has a a soft repulsion at smaller $r$, followed by a 
shallow minimum which supports the ground state at 
$125.51$ mK. As stated earlier, we fail to get the first excited state 
in this effective potential, by our numerical calculation. Next we 
calculate the isospectral potential $\hat{\omega}_0(r,\lambda)$, for 
chosen values of $\lambda$. 
We checked that for large values of 
$\lambda$, calculated $\hat{\omega}_0(r,\lambda)$ is practically 
indistinguishable from $\omega_0(r)$. For small values of $\lambda$, 
the isospectral potential develops a deep and narrow well near the 
origin (left side well, LSW) with a barrier (intermediate barrier, IB) 
separating it from the shallow well (right side well, RSW). As 
$\lambda$ decreases towards zero, LSW becomes deeper, 
narrower and closer to the origin and at the same time, IB becomes higher. 
\begin{center}
\begin{table}[!h]
\caption{ Parameters of the original and the isospectral potentials. $\lambda$ = $\infty$ corresponds to the original potential. LSW, IB and RSW respectively represent the left side well, the intermediate barrier and the right side well.
Position values are in a.u. and energies are in mK.}
\vskip 5pt
\begin{tabular} {|l|l|l|l|l|l|l|}
\hline
$\lambda$
& \multicolumn{2}{|l|}{LSW} & \multicolumn{2}{|l|}{IB} & \multicolumn{2}{|l|}{RSW} \\ \cline{2-7}
		&  $r_{min}^{left}$ &  $\omega_{min}^{left}$ & $r_{max}$ & $\omega_{max}$ & $r_{min}^{right}$ & $ \omega_{min}^{right}$ \\ \hline
$\infty$   &        &         &          &         & 8.999 & -3.238 \\ \hline
0.00010    & 4.607  & -1.897  & 5.327    & 8.333   & 9.087 & -2.848 \\ \hline
0.00005    & 4.527  & -8.119  & 5.137    & 10.23   & 9.087 & -2.847 \\ \hline
0.00002    & 4.457  & -22.133 & 4.943    & 13.874  & 9.085 & -2.846  \\ \hline
0.00001    & 4.417  & -40.254 & 4.817    & 18.019  & 9.085 & -2.846  \\ \hline
\end{tabular}
\end{table}
\end{center}
In Table~III, we present numerical values of the parameters of the 
original (corresponding to $\lambda=\infty$) and isospectral 
potentials $\hat{\omega}_0(r,\lambda)$ for $\lambda=0.00010,0.00005,0.00002$ and $0.00001$. For each isospectral potential, we give values of 
the position ($r^{left}_{min}$) and value ($\omega^{left}_{min}$) of 
the minimum of LSW, position ($r_{max}$) and value 
($\omega_{max}$) of the maximun of IB, 
and position ($r^{right}_{min}$) and value 
($\omega^{right}_{min}$) of the minimum of RSW. 
We have checked by numerical calculation that the energy of both the 
ground and the first excited state remains independent of the choice of 
$\lambda$, as theory predicts. However, 
very small values of $\lambda$ are not convenient for numerical 
calculation of the energy of the excited state, as numerical errors 
creep in, due to extreme narrowness of the LSW, in which the excited 
state resides. Thus a judicious choice of the 
value of $\lambda$ is necessary. By careful investigations for different 
values of $\lambda$, 
we find that the range $0.00002 - 0.00005$ for the value of $\lambda$ is 
optimum for minimizing errors. As two typical cases demostrating the 
behavior of isospectral potentials, we plot $\hat{\omega}_{0}(r,\lambda)$ 
for $\lambda=0.00002$ (blue, dotted curve)
and $\lambda=0.00005$ (green, dashed curve) in Fig.~1. The very weakly bound first excited 
state in the original shallow potential $\omega_0(r)$ now becomes 
strongly bound within the deep and narrow LSW of $\hat{\omega}_0(r,\lambda)$. 
The deep well and the adjacent high barrier strongly localizes this state 
in the isospectral potential. We solve Eq.~(11), with $\omega_0(r)$ 
replaced by $\hat{\omega}_0(r,\lambda)$, subject to appropriate boundary 
conditions to calculate the energy of the first excited state. 
Following this isospectral technique, we do indeed get a bound first excited state. 
We also verified that its energy is the same, within estimated numerical 
errors, for values of $\lambda$ lying within the chosen range. The binding energy 
of the first excited state of trimer is thus found to be $2.270$ mK. We present 
this result, together with results of other sophisticated calculations 
reported so far in Table~IV. \\
\begin{center}
\begin{table}[!h]
\caption{Results for energy of the first excited state of $^{4}$He-trimer (in units of mK) obtained 
by different methods. }
\begin{tabular} {|l|l|}
\hline
Reference  &  Energy \\ \hline
Ref [11]  &  -2.282 \\ \hline
Ref [7]  & -2.280  \\ \hline
Ref [8] & -2.277 \\ \hline
Present Method & -2.270 \\ \hline
\end{tabular}
\end{table}
\end{center}
Here we remark that the isospectral formalism is an efficient tool to calculate near-zero energy states in shallow potential which supports at least one bound state. 
In the supersymmetric isospectral formalism, one can in principle calculate the energy value as well as the wave function. In our earlier calculation of the resonances of halo nuclei, we used the wave function of the isospectral potential. We calculated the probability of the system to be trapped in the well-barrier combination which facilitates the accurate determination of resonance energy~\cite{tfr,xzz}. We have observed that the resonance energy is independent of $\lambda$. However in the present application to calculate the bound state wave function in the isospectral potentials and to calculate other physical observables we should take a different approach. From Fig.~1, we see that decreasing $\lambda$ gradually keeps the long-range part of the isospectral potentials almost unchanged whereas the short-range part has a drastic change. Naturally the wave functions of the isospectral family are now $\lambda$ dependent and obviously the calculated physical observables [e.g. average radius of the system] will be $\lambda$ dependent. In the SUSY isospectral formalism~\cite{ygt}, it is possible to calculate the wave function of the original potential $\omega_0(r)$ from the wave functions of the isospectral potential $\hat{\omega}_0(r,\lambda)$. 
The ground state $\hat{\zeta}_0(r,\lambda)$ and the first excited state $\hat{\zeta}_{1}(r,\lambda)$ of the isospectral potential $\hat{\omega}_0(r,\lambda)$ corresponding to the energies $E_0$ and $E_1$ respectively are related to the ground state $\zeta_0(r)$ and the first excited state $\zeta_1(r)$ of the original potential $\omega_0(r)$ by~\cite{ygt}
\begin{equation}
 \hat{\zeta}_0(r,\lambda) = \frac{\zeta_0(r)}{I_0+\lambda}
\end{equation}
and \begin{equation}
     \hat{\zeta}_{1}(r,\lambda)= (E_1-E_0)\zeta_1+\hat{\zeta}_0{\mathcal W}(\zeta_0,\zeta_1)
    \end{equation}
where ${\mathcal W}(\zeta_0,\zeta_1)= \zeta_0\zeta_1^{\prime}-\zeta_1\zeta_0^{\prime}$ is the Wronskian and $\zeta_0(r)$ and $\zeta_1(r)$ are well-behaved functions. We can calculate the ground state wave function $\hat{\zeta}_0(r,\lambda)$ of the isospectral potential $\hat{\omega}_0(r,\lambda)$ from the original gound state wave function $\zeta_0(r)$ (which is known as we calculated it to construct the family of isospectral potentials) using Eq.~(19). Alternatively we can obtain $\hat{\zeta}_0(r,\lambda)$ directly by solving for the isospectral potential $\hat{\omega}_0(r,\lambda)$. Also we can obtain $\hat{\zeta}_1(r,\lambda)$ by solving for the isospectral potential $\hat{\omega}_0(r,\lambda)$ with energy $E_1$. Thus Eq.~(20) is a first order differential equation in $\zeta_1(r)$ and we can solve it subject to proper boundary conditions $\zeta_1(0)=0$ and $\zeta_1^{\prime}(0) \neq 0$ to obtain $\zeta_1(r)$. With these wave functions we can calculate any physical properties of the system and this time they will be independent of the $\lambda$. However usually in the SUSY isospectral formalism, people are interested only in the bound state energy and in this paper we are also interested only in the energy of the elusive first excited state.\\

Next we discuss in detail about the choice of $\lambda$ parameter. As in our earlier works~\cite{xzz,ffe,fcr}, 
we observe that the energy is independent of the parameter $\lambda$. But making $\lambda$ too small, may create large numerical error in the numerically calculated wave function $\hat{\zeta}_1(r,\lambda)$, as the 
well becomes extremely narrow and very deep. Then the wave function changes very rapidly over a very small interval. Clearly its derivatives are inaccurate, which in turn affects the accuracy of the wave function at the next mesh point. The errors accumulate. As we solve the 
isospectral potential numerically, large numerical error will result for very small $\lambda$. This may mask the overall accuracy. 
Thus the accuracy of the results are very crucially dependent on the choice of $\lambda$. There is no prescribed rule to choose $\lambda$. It is in general 
chosen by trial. One can choose finer mesh interval within the LSW for a small $\lambda$. But the cumulative error at successive mesh point, increases as the total number of mesh points increases. Using the error estimate of the integration procedure (we use Runga-Kutta algorithm) we optimize the values of $\lambda$ and mesh interval to minimize the error. 
In our earlier calculations~\cite{xzz,ffe} for $2_{1}^{+}$ state of $^{6}$He, the shallowness of the effective potential is removed for $\lambda$ = 10, whereas $\lambda$ = 0.1 was required for $2_{2}^{+}$ state of the same system to get the expected behavior of the isospectral potential. Thus the optimum choice of $\lambda$ depends on the choice of the system and its state. \\

Now in principle the ground state energy can also be recalculated using the isospectral potentials, in which the ground state of the 
original potential now lies in the very deep well of the isospectrals. But, when we solve the isospectral potential numerically, large error may creep due to the 
extreme narrowness of the well, as discussed above. Thus when we use the isospectral formalism for the real many-body problems, ground state energy is more accurate when determined 
by the original potential and the excited states and resonance states are more accurate when we solve the isospectral potentials.\\ 

\section{Conclusions}
In conclusion, we remark that the isospectral potential with a judiciously chosen 
value of $\lambda$ can be very useful for an accurate calculation of near-zero 
energy states in a shallow potential and also the resonance states of halo and highly unstable systems. The present application to the very weakly bound and 
highly elusive first excited state in $^{4}$He trimer demonstrates the novelty 
and practical utility of this technique.\\

This work has been partially supported by FAPESP (Brazil), CNPq (Brazil), Department of Science and 
Technology (DST, India) and Department of Atomic Energy (DAE, India). 
B.C. wishes to thank FAPESP (Brazil) for providing financial assistance for her visit to the 
Universidade de S\~ao Paulo, Brazil, where part of this work was done. SKH wishes to thank the Council of Scientific and Industrial Research (CSIR), India for a Junior Research Fellowship. T.K.D. acknowledges the University Grants Commission (UGC, India) 
for the Emeritus Fellowship.\\


\begin{thebibliography}{References}
\bibitem{uhi} V. Efimov, {Phys. Lett. B {\bf{33}}, 563 (1970)}.
\bibitem{iju} V. Efimov, {Nucl. Phys. A {\bf{210}}, 157 (1973)}.
\bibitem{KKZ} T. Kraemar {\it et al}, Nature {\bf {440}}, 315
(2006); S. Knoop {\it et al}, Nature Phys. Lett. {\bf {5}},
227 (2009); M. Zaccanti {\it et al}, Nature Phys. {\bf {5}},
586 (2009).
\bibitem{gft} T. Cornelius and W. Gl\"ockle, {J. Chem. Phys. {\bf{85}}, 3906 (1986)}.
\bibitem{tgy} B. D. Esry, C. D. Lin, and C. H. Greene, {Phys. Rev. A {\bf{54}}, 394 ( 1996)}.
\bibitem{esw} E. Nielsen, D. V. Fedorov and A. S. Jensen, 
{J. Phys. B {\bf{31}}, 4085 (1998)}
\bibitem{sad} V. Roudnev and S. Yakovlev, {Chem. Phys. Lett. {\bf{328}}, 97 (2000)}.
\bibitem{yhu} P. Barletta and A. Kievsky, {Phys. Rev. A {\bf{64}}, 042514 (2001)}.
\bibitem{ijj} W. Sandhas, E. A. Kolganova, Y. K. Ho and A. K. Motovilov, {Few-Body Syst. {\bf{34}}, 137 (2004)}.
\bibitem{uuu} T. Gonz\'alez-Lezana {\it {et al.}}, {Phys. Rev. Lett. {\bf{82}}, 1648 (1999)}.
\bibitem{szx} A. K. Motovilov, W. Sandhas, S. A. Sofianos and E. A. Kolganova, {Eur. Phys. J. D {\bf{13}}, 33 (2001)}.
\bibitem{swa} H. Suno and B. D. Esry, {Phys. Rev. A {\bf{ 78}}, 062701 (2008)}.
\bibitem{gff} E. Braaten, H. W. Hammer, D. Kang, and L. Platter, {Phys. Rev. A {\bf{78}}, 043605 (2008)}.
\bibitem{ews} W. Sch\"ollkopf and J. P. Toennies, {Science {\bf{266}}, 1345 (1994)}.
\bibitem{ytg} R. Br\"uhl {\it {et al.}}, {Phys. Rev. Lett. {\bf{95}}, 063002 (2005)}.
\bibitem{fabre} M. Fabre de la Ripelle, {Ann. Phys. (N.Y.) {\bf{147}}, 281 (1983)}.
\bibitem{hhh} T. K. Das, B. Chakrabarti and S. Canuto, {J. Chem. Phys. {\bf{134}}, 164106 (2011)}.
\bibitem{ygt} F. Cooper, A. Khare and U. Sukhatme, {Phys. Rep. {\bf{251}}, 267 (1995)}.
\bibitem{rft} T. K. Das and B. Chakrabarti, {Phys. Rev. A {\bf{70}}, 063601 (2004)}.
\bibitem{edd} T. K. Das {\it{et al.}},{Phys. Lett. A {\bf{373}}, 258 (2009)}.
\bibitem{arj} A. R. Janjen and R. A. Aziz, {J. Chem. Phys. {\bf{103}}, 9626 (1995)}.
\bibitem{fdr} K. T. Tang, J. P. Toennies, C. L. Yiu, {Phys. Rev. Lett. {\bf{74}}, 1546 (1995)}.
\bibitem{raa} R. A. Aziz and M. J. Slaman, {J. Chem. Phys. {\bf 94}, 8047 (1991)}.
\bibitem{raaz} R. A. Aziz, V. P. S. Nain, J. S. Carley, W. L. Taylor, 
and G. T. McConville, {J. Chem. phys. {\bf 70}, 4330 (1979)}.
\bibitem{edr} T. K. Das, H. T. Coelho and M. Fabre de la Ripelle, {Phys. Rev. C {\bf{26}}, 2281 (1982)}.
\bibitem{trf} M. Lewerenz, {J. Chem. Phys. {\bf{106}}, 4596 (1997)}.
\bibitem{gho} D. Bressanini {\it et al.} {J. Chem. Phys. {\bf{112}}, 717 (2000)}.
\bibitem{kli} D. Blume and C. H. Greene, {J. Chem. Phys. {\bf{112}}, 8053 (2000)}.
\bibitem{fvr} R. Grisenti, W. Sch\"ollkopf and J. P. Toennies, {Phys. Rev. Lett. {\bf{85}}, 2284 (2000)}.
\bibitem{tfr} T. K. Das and B. Chakrabarti, {Phys. Letts A {\bf{288}}, 4 (2001)}.
\bibitem{xzz} S. K. Dutta, T. K. Das, M. A. Khan and B. Chakrabarti {Few-Body Systems {\bf{35}}, 33 (2004)}.
\bibitem{ffe} S. K. Dutta, T. K. Das, M. A. Khan and B. Chakrabarti, {J. Phys. G: Nucl. Part. Phys. {\bf{29}}, 2411 (2003)}. 
\bibitem{fcr} B. Chakrabarti, {Proceedings of Conference on 'Non-Hermitian Hamiltonian in Quantum Physics', BARC, India, June 13 - 16, 2009 [Pramana, Journal of Physics, {\bf{73}}, 405 (2009)]}.

\end{thebibliography}
\end{document}